\documentstyle[preprint,aps,epsf]{revtex}
\begin{document}
\preprint{hep-th/9510141, CU-TP-712, SNUTP-95-106}
\draft
\title{ Anyonic Bogomol'nyi Solitons in \\ a Gauged $O(3)$ Sigma Model}
\author{Kyoungtae Kimm${}^{(1)}$, Kimyeong Lee${}^{(2)}$,
               and Taejin Lee${}^{(3)}$}
\address{
{${}^{(1)}$Department of Physics, Seoul National University, Seoul
151-741, Korea} \\ {${}^{(2)}$ Physics Department, Columbia
University, New York, N.Y. 10027, U.S.A}
\\ {${}^{(3)}$ Department of
Physics, Kangwon National University, Chuncheon 200-701, Korea} }
\maketitle
\begin{abstract}

We introduce the self-dual abelian gauged $O(3)$ sigma models where
the Maxwell and Chern-Simons terms constitute the kinetic terms for
the gauge field.  These models have quite rich structures and various
limits. Our models are found to exhibit both symmetric and broken
phases of the gauge group.  We discuss the pure Chern-Simons limit in
some detail and study rotationally symmetric solitons.

\end{abstract}

Recently the gauged $O(3)$ sigma models\cite{gaugeo3-1,gaugeo3-2} have
been studied in three dimensions with the Maxwell term where the gauge
group $U(1)$ is a subgroup of $O(3)$.  These models differ from the
gauged $O(3)$ sigma model previously discussed\cite{wilczek,kara,tlee}
in that the gauge field is coupled to the scalar fields through a
$U(1)$ electric current rather than the topological one.  If a
specific potential is chosen, then the Bogomol'nyi type energy bound
is found to exist.  This bound is shown to be saturated by topological
lumps carrying nonzero magnetic flux.  The size of these topological
lumps depends on the magnetic flux but their mass is independent of
the magnetic flux.  Subsequently, the self-dual gauged sigma models
with both Maxwell and Chern-Simons terms have been
discussed\cite{Ghosh}, where both topological and nontopological
solitons are found.

However, the vacua of all above models respect the gauge symmetry, so
the solitons live in the symmetric phase. In this paper, we generalize
these models such that the asymmetric phase where the gauge symmetry
is spontaneously broken is also admitted. For the generality, we may
also include a uniform external charge density.  It seems that the
structures of these models are rather rich.

The models exhibit two inequivalent symmetric phases and one
asymmetric phase, depending on a parameter of the theory. By studying
the rotationally symmetric self-dual solitons, we found that in the
symmetric phases there are nontopological $Q$-lumps\cite{Leese} with
or without vortices at their centers.  Also there are topological
lumps carrying nonzero unquantized magnetic flux.  In general their
mass and size depend on their magnetic flux.  In the broken phase,
there are two types of topological vortices of quantized magnetic
flux.  Curiously, it seems in the asymmetric phase that there exist
topological lumps which are not rotationally symmetric.

Let us start with a scalar field ${\mbox{\boldmath $\phi$}}(x)$ which
is a map from the 3 dimensional spacetime to the two-sphere of unit
radius.  As we have scaled ${\mbox{\boldmath $\phi$}}$ to be a unit
vector, the spatial coordinates and the coupling parameters will be
dimensionless.  For a given ${\mbox{\boldmath $\phi$}}(x)$ field
configuration, one can construct a topological current\cite{belavin}
\begin{eqnarray}
k_{\alpha} =\frac{1}{8\pi} \epsilon_{\alpha\beta\gamma}
{\mbox{\boldmath $\phi$}}\!\cdot\!\partial^{\beta}{\mbox{\boldmath
$\phi$}} \times \partial^{\gamma}{\mbox{\boldmath $\phi$}},
\end{eqnarray}
which is conserved trivially.  If ${\mbox{\boldmath $\phi$}}$
approaches a constant unit vector at the spatial infinity, we can
compactify two dimensional space as a unit sphere and regard
${\mbox{\boldmath $\phi$}}$ as a mapping from a two-sphere to a
two-sphere. The integer-valued associated degree in this case is given
by the charge $S = \int d^2 \! x k_0$ of the topological current.

We introduce the $U(1)$ gauge coupling by a covariant derivative
\begin{eqnarray}
D_{\alpha} {\mbox{\boldmath $\phi$}} =
\partial_{\alpha}{\mbox{\boldmath $\phi$}} + A_{\alpha}
{\mbox{\boldmath $n$}} \times {\mbox{\boldmath $\phi$}} ,
\end{eqnarray}
where ${\mbox{\boldmath $n$}}= (0,0,1)$ is a unit vector.  The $U(1)$
gauge group is a subgroup of the $O(3)$ rotational group of
${\mbox{\boldmath $\phi$}}(x)$.  The gauge invariant generalization of
the topological current $k_\alpha$ is
\begin{eqnarray}
K_{\alpha} = \frac {1}{8\pi} \epsilon_{\alpha \beta \gamma}
\left( {\mbox{\boldmath $\phi$}}\!\cdot\!
D^{\beta}{\mbox{\boldmath $\phi$}}\times D^{\gamma}{\mbox{\boldmath
$\phi$}} + F^{\beta\gamma} (v - {\mbox{\boldmath $n$}} \!\cdot\!
{\mbox{\boldmath $\phi$}}) \right),
\end{eqnarray}
where $v$ is a free real parameter. This current differs from
$k_{\alpha}$ only by the curl of a vector field:
\begin{eqnarray}
K_{\alpha} = k_{\alpha} + \frac{1}{4\pi} \epsilon_{\alpha \beta
\gamma} \partial^{\beta} \left((v -{\mbox{\boldmath$n$}}\!\cdot\!
{\mbox{\boldmath $\phi$}}) A^{\gamma} \right).
\end{eqnarray}
The corresponding conserved topological charge is $T = \int d^2x K^0$
which may differ from the degree $S$.

As we will see, we can impose various boundary conditions at spatial
infinity on the finite energy configurations.  Depending upon the
boundary conditions, the gauge symmetry can be realized in the
symmetric phase or the asymmetric one: (1) For the symmetric phases
\begin{eqnarray}
\lim_{ |{\mbox{\boldmath$x$}}| \rightarrow \infty}
       {\mbox{\boldmath $\phi$}}(t,{\mbox{\boldmath$x$}}) = \pm
{\mbox{\boldmath $n$}} ,
\end{eqnarray}
where the gauge boson is massless; (2) If $|v| <1$, one can impose
\begin{eqnarray}
\lim_{|{\mbox{\boldmath$x$}}| \rightarrow \infty} {\mbox{\boldmath $n$}}
         \!\cdot\! {\mbox{\boldmath $\phi$}}(t,{\mbox{\boldmath$x$}})
= v .
\end{eqnarray}
and the asymmetric phase is realized.  In the symmetric phase, the
mapping ${\mbox{\boldmath $\phi$}}: R^2 \rightarrow S^2$ can be
regarded as a map from a two-sphere to another two-sphere, and $S$
measures the associated second homotopy which takes an integer value.
However, in the asymmetric phase, the mapping ${\mbox{\boldmath
$\phi$}}$ is not necessarily such a map.  The vacuum manifold is a
unit circle and ${\mbox{\boldmath $\phi$}}(t,r=\infty,\theta)$ in the
polar coordinate $(r,\theta)$ of the spatial plane is a map from $S^1
\rightarrow S^1$.  If the mapping at the spatial infinity has any
nontrivial first homotopy, the mapping ${\mbox{\boldmath $\phi$}}: R^2
\rightarrow S^2$ covers the two-sphere partially and the degree $S$
becomes fractional.  On the other hand, if the mapping at the spatial
infinity has the trivial homotopy, the degree $S$ will be an integer.

In general the kinetic term for the gauge field in three dimensions
consists of the Maxwell term and the Chern-Simons term and a uniform
external charge density may be coupled to the gauge
field\cite{Klee,LeeYi}.  In order to make the model self-dual, we also
need a neutral scalar field $N$ which couples to the ${\mbox{\boldmath
$\phi$}}$ field.  Hence the general Lagrangian we consider here is
given by
\begin{eqnarray}
{\cal L}_1 = -\frac{1}{4e^2}(F_{\alpha\beta})^2 +
\frac{\kappa}{2}\epsilon^{\alpha\beta\gamma} A_\alpha \partial_\beta
A_\gamma + \frac{1}{2e^2}(\partial_\alpha N)^2 +\frac{1}{2}(D_\alpha
{\mbox{\boldmath $\phi$}})^2 - U({\mbox{\boldmath $\phi$}},N) - \rho_e
A_0 ,
\end{eqnarray}
where $e^2,\kappa, \rho_e$ are dimensionless parameters. By a similar
procedure as in Ref. \cite{Bogo} we can find the potential $U$ with
which the Bogomol'nyi type bound on the energy functional exists.  The
potential are found to be one of two special potentials:
\begin{eqnarray}
U({\mbox{\boldmath $\phi$}}, N)_\pm = \frac{e^2}{2}(\kappa N + (v-
{\mbox{\boldmath $n$}} \!\cdot\! {\mbox{\boldmath $\phi$}})) ^2 +
\frac{1}{2} N^2 ({\mbox{\boldmath $n$}} \!\times\! {\mbox{\boldmath
$\phi$}})^2 \mp \rho_e N .
\end{eqnarray}
Then, the energy functional is bounded by $\pm 4\pi T$ depending on
the sign of the potential.

When $\rho_e \ne 0$, this self-dual model seems to have very rich
structures similar to those considered in Ref. \cite{LeeYi} and we
hope to pursue them elsewhere.  Here we restrict ourselves to the case
$\rho_e=0$ for the sake of simplicity.  In the pure Maxwell limit
$\kappa \rightarrow 0$, the potential becomes
\begin{eqnarray}
U_{m} = \frac{e^2}{2}(v -{\mbox{\boldmath $n$}} \!\cdot\!
{\mbox{\boldmath $\phi$}})^2 + \frac{1}{2}N^2 ({\mbox{\boldmath $n$}}
\!\times\!  {\mbox{\boldmath $\phi$}})^2
\end{eqnarray}
The model considered in Ref. \cite{gaugeo3-1} corresponds to case when
$v=1$.  On the other hand, in the pure Chern-Simons limit, $e^2
\rightarrow \infty$, classically $N = - (v-{\mbox{\boldmath $n$}}
\!\cdot\!  {\mbox{\boldmath $\phi$}})/\kappa$ and the potential
becomes
\begin{eqnarray}
 U_{cs} = \frac{1}{2\kappa^2} (v-{\mbox{\boldmath $n$}} \!\cdot\!
{\mbox{\boldmath $\phi$}})^2({\mbox{\boldmath $n$}} \!\times\!
{\mbox{\boldmath $\phi$}})^2 .
\end{eqnarray}
This model with $v=1$ is the one considered in Ref. \cite{Ghosh}.

In this paper we focus our attention on the pure Chern-Simons case
where the Lagrangian is given by
\begin{eqnarray}
 {\cal L}_{cs} = \frac{\kappa}{2}\epsilon^{\alpha\beta\gamma} A_\alpha
\partial_\beta A_\gamma + \frac{1}{2}(D_\alpha {\mbox{\boldmath
$\phi$}})^2 -
\frac{1}{2\kappa^2} (v-{\mbox{\boldmath $n$}} \!\cdot\!
{\mbox{\boldmath $\phi$}})^2({\mbox{\boldmath $n$}} \!\times\!
{\mbox{\boldmath $\phi$}})^2 .
\end{eqnarray}
When $|{\mbox{\boldmath $\phi$}}-{\mbox{\boldmath $n$}}| \ll 1$ and
$0<1-v \ll 1$, the above model reduces to the self-dual Chern-Simons
Higgs systems\cite{Hong,jlw}. When $|v|<1$, there are three degenerate
minima: (1) the two symmetric phases where ${\mbox{\boldmath $\phi$}}=
\pm {\mbox{\boldmath $n$}}$ and the mass of charged scalar particles
is $m_\pm = |(v \pm 1)/\kappa| $.  (2) the broken phase where
${\mbox{\boldmath $n$}}\!\cdot\!{\mbox{\boldmath $\phi$}}=v$ and the
masses of neutral vector and scalar particles are given by $|( 1 - v^2
)/ \kappa |$. When $|v| \ge 1$, only the symmetric phases remain.

The Gauss law obtained from the variation of $A_0$ is
\begin{eqnarray}
\label{eq:Gauss}
 \kappa F_{12} + {\mbox{\boldmath $n$}} \!\cdot\!  {\mbox{\boldmath
$\phi$}} \!\times\! D_0 {\mbox{\boldmath $\phi$}} = 0 ,
\end{eqnarray}
and the gauged $U(1)$ current is given by $J^\alpha = {\mbox{\boldmath
$n$}} \!\cdot\! {\mbox{\boldmath $\phi$}}
\!\times\! D^\alpha {\mbox{\boldmath $\phi$}}$.
The Gauss law implies that the total magnetic flux $\Psi = \int d^2x
F_{12}$ is related to the total charge $Q = \int d^2x J^0$ by the
equation $ \kappa \Psi = - Q$.

The conserved energy functional is
\begin{eqnarray}
E = \frac{1}{2} \int \, d^2 x \, \left( (D_0 {\mbox{\boldmath
$\phi$}})^2 + (D_i {\mbox{\boldmath $\phi$}})^2 + U_{cs} \right).
\end{eqnarray}
The energy functional can be rewritten as
\begin{eqnarray}
E = \frac{1}{2} \int d^2 x\, \left( (D_0 {\mbox{\boldmath $\phi$}} \pm
\frac{1}{\kappa} (v -{\mbox{\boldmath $n$}}\!\cdot\!  {\mbox{\boldmath
$\phi$}}){\mbox{\boldmath $n$}}
\!\times\! {\mbox{\boldmath $\phi$}} )^2  +
( D_1 {\mbox{\boldmath $\phi$}} \pm {\mbox{\boldmath $\phi$}}
\times  D_2 {\mbox{\boldmath $\phi$}} )^2 \right) \pm
4\pi T ,
\end{eqnarray}
where we take advantage of ${\mbox{\boldmath $\phi$}} \!\cdot\!
D_\alpha {\mbox{\boldmath $\phi$}}=0$ and the Gauss law. Here we also
discarded a total derivative whose contribution vanishes.  Thus, there
is a bound on the energy,
\begin{eqnarray}
 E \ge 4\pi |T| ,
\end{eqnarray}
and this bound is saturated if the self-dual equations hold:
\begin{eqnarray}
\label{Bogeq}
D_1{\mbox{\boldmath $\phi$}} &=& \mp {\mbox{\boldmath $\phi$}}
\times D_2{\mbox{\boldmath $\phi$}} , \\
D_0 {\mbox{\boldmath $\phi$}} &=& \pm \frac{1}{\kappa^2}(v
-{\mbox{\boldmath $n$}} \!\cdot\!  {\mbox{\boldmath
$\phi$}})({\mbox{\boldmath $n$}} \!\times\!  {\mbox{\boldmath
$\phi$}}).\label{eq:selfdual2}
\end{eqnarray}
The second equation yields ${\mbox{\boldmath $n$}} \!\cdot\!
\partial_0{\mbox{\boldmath $\phi$}}=0$,
which implies the energy density is static in time. We can choose the
gauge such that the field configurations satisfying the self-dual
equations are static.  Putting Eqs. (\ref{eq:Gauss}) and
(\ref{eq:selfdual2}) together, we find
\begin{eqnarray}
F_{12} &=& \pm \frac{1}{\kappa^2}(v -{\mbox{\boldmath $n$}} \!\cdot\!
{\mbox{\boldmath $\phi$}})({\mbox{\boldmath $n$}} \!\times\!
{\mbox{\boldmath $\phi$}})^2.
\end{eqnarray}
The conserved angular momentum is given by
\begin{eqnarray}
J = - \int d^2x \, \epsilon_{ij} x^i D_0{\mbox{\boldmath $\phi$}}
\!\cdot\!  D_j {\mbox{\boldmath $\phi$}}.
\end{eqnarray}
As we will see, the solitons in this system carry in general
fractional angular momenta, thus can be regarded as anyons.

In the parameterization
\begin{eqnarray}
\label{hedge}
{\mbox{\boldmath $\phi$}}(x) = (\sin f(x) \cos \psi(x) , \sin f(x)
\sin \psi(x) , \cos f (x)),
\end{eqnarray}
we may rewrite the conserved angular momentum as
\begin{eqnarray}
 J = \int d^2 x \,\epsilon_{ij} x^i \biggl( \partial_0 f \partial_j f
- \kappa (\partial_j \psi + A_j) F_{12} \biggr).
\end{eqnarray}
if we make use of the Gauss law.  In terms of this parameterization
the self-dual equations become
\begin{eqnarray}
\partial_i f = \pm \epsilon_{ij} (\partial_j \psi + A_j) \sin f ,\\
\kappa^2 F_{12} = \pm (v-\cos f)\sin^2 f .
\end{eqnarray}
{}From these equations one can deduce that there may be two kind of
vorticities, of which asymptotics are $f=0$ and $f=\pi$.

Let us now consider rotationally symmetric solitons satisfying the
self-dual equations.  In the polar coordinate $(r,\theta)$, let us
take an ansatz $f= f(r)$, $\psi= N\theta$ with an integer $N$ and
$A_\theta = a(r) $ to respect the rotational symmetry.  We may choose
the upper sign in the above self-dual equations without lose of
generality, assuming that the degree $T$ is positive.  The self-dual
equations become
\begin{eqnarray}
    r f'(r) = ( N + a(r) ) \sin f(r) ,\label{eq:sd1} \\ a'(r) =
\frac{r}{\kappa^2 } (v - \cos f(r)) \sin^2 f(r).\label{eq:sd2}
\end{eqnarray}
With $\alpha \equiv a(\infty)$ the degree for this ansatz is given by
\begin{eqnarray}
 T = \frac{N}{2} \left[ \cos f(0) - \cos f(\infty)\right] +
\frac{\alpha}{2} \left[v-\cos f(\infty)\right],
\end{eqnarray}
while the magnetic flux and the angular momentum of the solution are
$\Psi = 2\pi \alpha $ and $J = \pi \kappa ((N+ \alpha)^2 - N^2)$.

Eqs.(\ref{eq:sd1}) and (\ref{eq:sd2}) tell the behavior of the
solution near $r=0$, which is rather straightforward.  For $A_i$ to be
nonsingular at the origin, $a(0) = 0 $. When $f(0)\ne 0, \, \pi$,
${\mbox{\boldmath $\phi$}}$ is nonsingular only if $N=0$.  When
$f(0)=0$, $N$ should be positive and $f(r)\propto r^N$ near $r=0$.
When $f(0) = \pi$, $N$ should be negative and $f(r)\propto r^{|N|}$
near $ r = 0$.

The behavior of the solution near $r=\infty$ depends on whether
$f(\infty) = 0 , f_v$ or $ \pi$: When $f(\infty)=0$, $f(r) \propto
r^{N+\alpha}$ and $a(r) -\alpha \propto r^{2(N+\alpha +1)}$ for large
$r$. In this case the asymptotic behavior is consistent if $\alpha <
-1-N $. When $f(\infty)=\pi$, $\pi - f(r) \propto r^{-N-\alpha}$ and
$a(r)-\alpha
\propto r^{-2N-2\alpha +2}$. Thus, the consistent
asymptotic behavior requires that $ \alpha> 1-N$.  When $f(\infty) =
f_v$, $\alpha = -N$ and $f(r),a(r)$ approach to their asymptotic
values exponentially.

{}From Eq.(\ref{eq:sd1}) it follows that the range of the function
$f(r)$ lies between $0$ and $\pi$. We can show this by the
contradiction.  Let us assume $f(r)$ to cross zero near $r=r_0$.
{}From Eq.(\ref{eq:sd1}), we see that near $r=r_0$, $ f \propto
e^{(N+a(r_0))(r-r_0)/r_0} $, which cannot vanish. Thus, $f(r)>0$.
Also the same procedure shows that it cannot cross $\pi$ in the range
$(0,\infty)$, i.e., $f(r)< \pi$.

Now we call our attention to the solutions when $|v|<1$ and discuss
the results of the numerical analysis.  Then we will consider later
the cases of $|v| \ge 1$ where the solutions for $|v|<1$ become
degenerate.

\vskip 2.0em

\noindent {\bf a)} When $  f(0)\neq 0,\,\pi$ and $N= 0$.

\vskip 1.5em
We define $0<f_v<\pi $ such that $\cos f_v = v$.  {}From
Eqs.(\ref{eq:sd1}) and (\ref{eq:sd2}) we see that if $f(0)>f_v$, both
$f(r)$ and $a(r)$ are increasing functions. At spatial infinity, it
should be that $f( \infty) = \pi $ and $\alpha>1$. The solution has
the mass $2\pi (v+1) \alpha$ and the angular momentum $J=\pi \kappa
\alpha^2$.  This is a nontopological soliton in the symmetric phase
$f=\pi$ without any vortex in its center.  If $f(0)<f_v$, we see that
$f(r)$ and $a(r)$ are decreasing function with $f(\infty)=0$ and $
\alpha<-1$. Its mass and spin are given by
$2\pi (1-v)(-\alpha)$ and $J=\pi \kappa \alpha^2$ respectively.  It is
again the nontopological soliton in the symmetric phase $f=0$ without
any vortex in its center.  Note that the charged particles in the
symmetric phases $f=0,\pi$ have the same mass per charge ratio as
these nontopological solitons, so they are marginally stable. Fig.1
depicts $f(r)/\pi$ for the cases $f_v =\pi/3$ with two different
choices for the values of $f(0)$.  Fig.2 shows $a(r)$ for the same
intial values.
\vskip 1.5em
\centerline{\epsfysize=5cm\epsffile{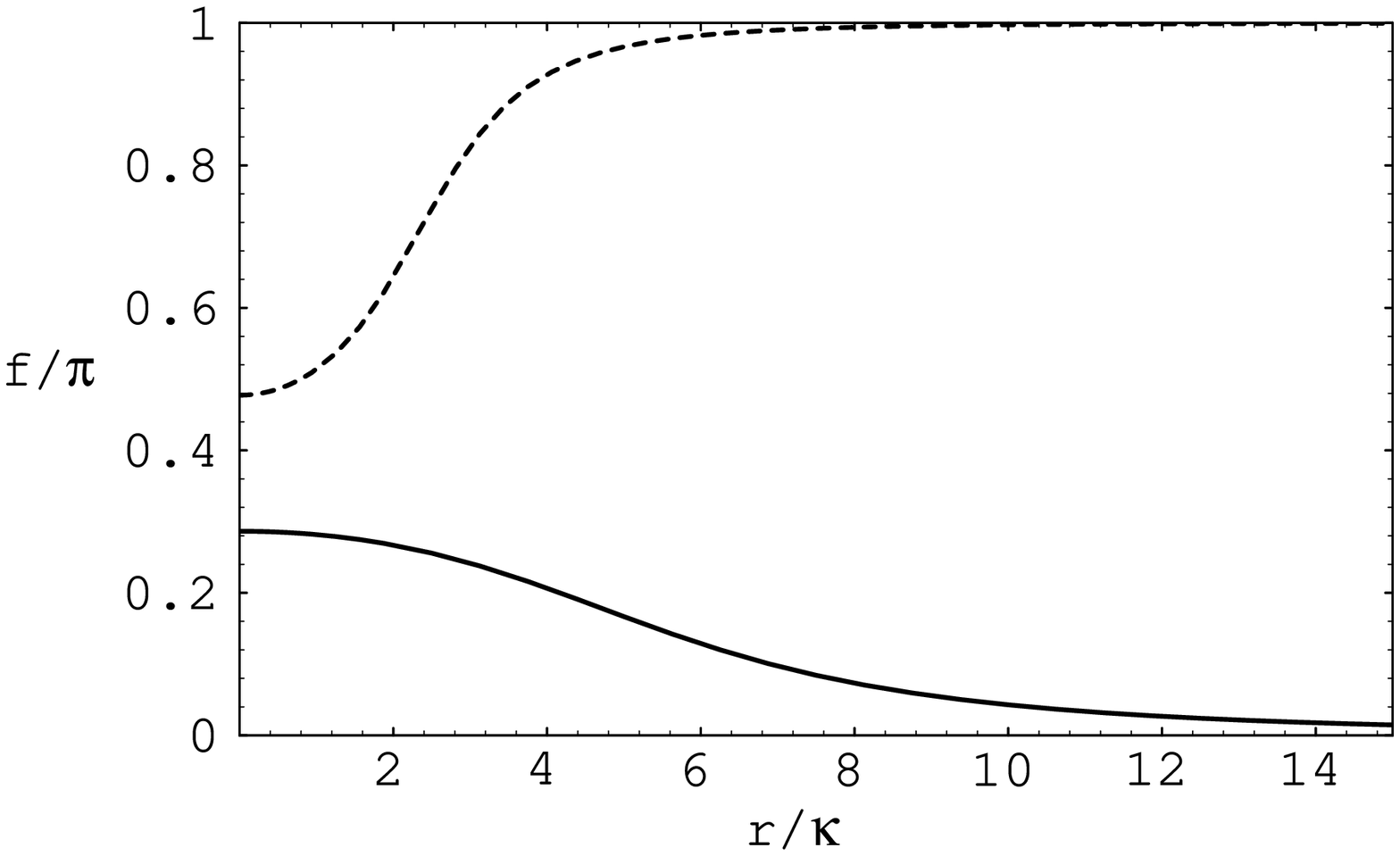}}
\hangindent\parindent{{\bf Fig.1}
Plotting $f(r)$ for $N=0$ and $f(0)\ne 0$ and $f_v=\pi/3$.  The dotted
line is for $f(0)=1.5$ and the solid line for $f(0)=0.9$.}
\vskip 1.5em
\centerline{\epsfysize=5cm\epsffile{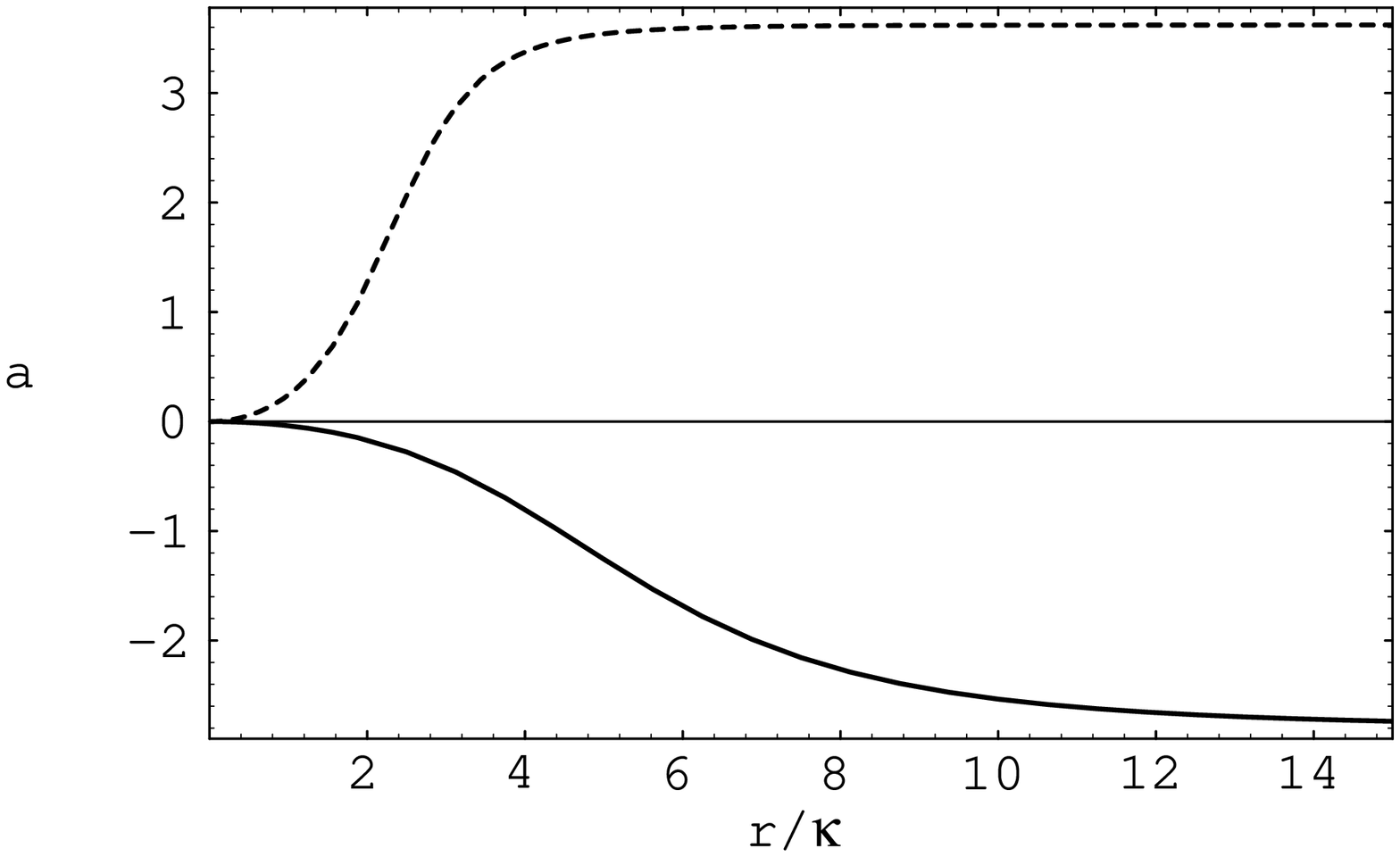}}
\hangindent\parindent{{\bf Fig.2}
Plotting $a(r)$ for $N=0$, $f(0)\ne 0$ and $f_v=\pi/3$.  The dotted
line is for $f(0)=1.5$ and the solid line for $f(0)=0.9$
respectively.}

\vskip 2em

\noindent {\bf b)} When $f(0)=0 $ and  $N \ge 1 $.

\vskip 1.5em

Let us now consider the case $f(0)=0$. There are three kinds of
solutions depending on whether $f(\infty) = 0, f_v$ or $\pi$. If
$f(\infty) = \pi$, we see that $ \alpha >- N+1$.  The function $f(r)$
is a monotonously increasing function, while $a(r)$ decreases and then
increases as $r$ increases, approaching to the value $\alpha$.  One
can see that the minimum of $a(r)$ is larger than $-N$.  Its mass and
angular momentum are $4\pi N + 2\pi(v+1)\alpha$, $J=\pi \kappa^2
\alpha(\alpha + 2 N)$ respectively.  It is a topological lump in the
symmetric phase $f=\pi$ with a nonzero degree $S= N$.

When $f(\infty) = f_v$, $f(r)$ is monotonically increasing while
$a(r)$ is monotonically decreasing with $a(\infty) = - N$. It is a
vortex in the asymmetric phase with magnetic flux $-2\pi N$ and mass
$2\pi N(1-v)$. Its spin is $-\pi \kappa N^2$.

When $f(\infty) = 0$, we see $a(r)$ is monotonically decreasing to $
\alpha<-(N+1)$. Here  $f(r)$ increases and then decreases, approaching
to zero.  This is a nontopological soliton with a vortex in the
middle.  Fig.3 describes the function $f(r)$ for these three cases
with $N=1, f_v=\pi/3$ in terms of the value $\alpha$. Fig.4 shows
$a(r)$ for the same values of $\alpha$.
\vskip 1.5em
\centerline{\epsfysize=5cm\epsffile{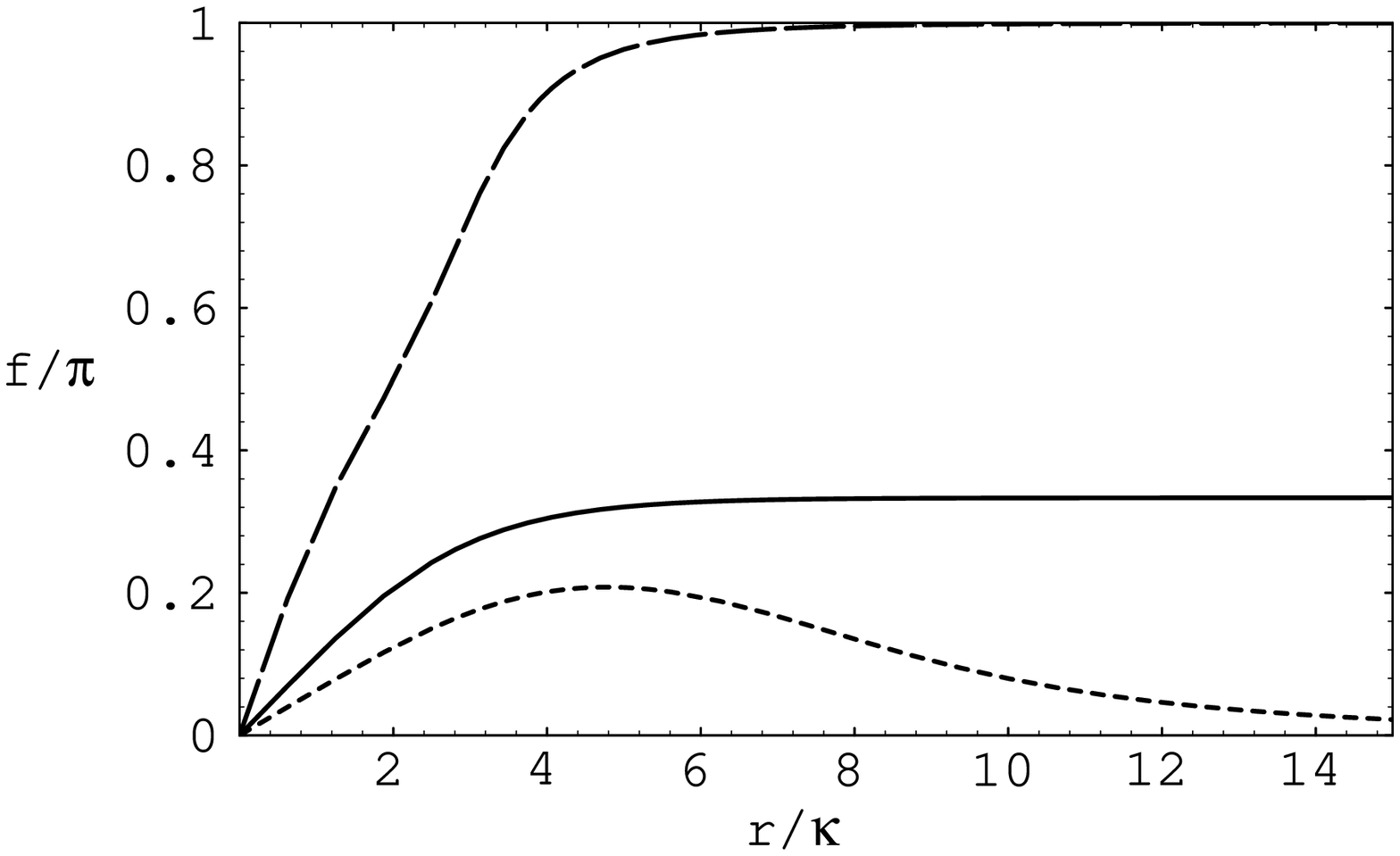}}
\hangindent\parindent{{\bf Fig.3}
Plotting $f(r)$ for $N=1$, $f(0)=0$ and $f_v=\pi/3$.  The dashed line
corresponds to $\alpha=3.53325$, the solid line to $\alpha=-1.0$ and
the dotted line to $\alpha=-4.50025$ respectively.}
\vskip 1.5em
\centerline{\epsfysize=5cm\epsffile{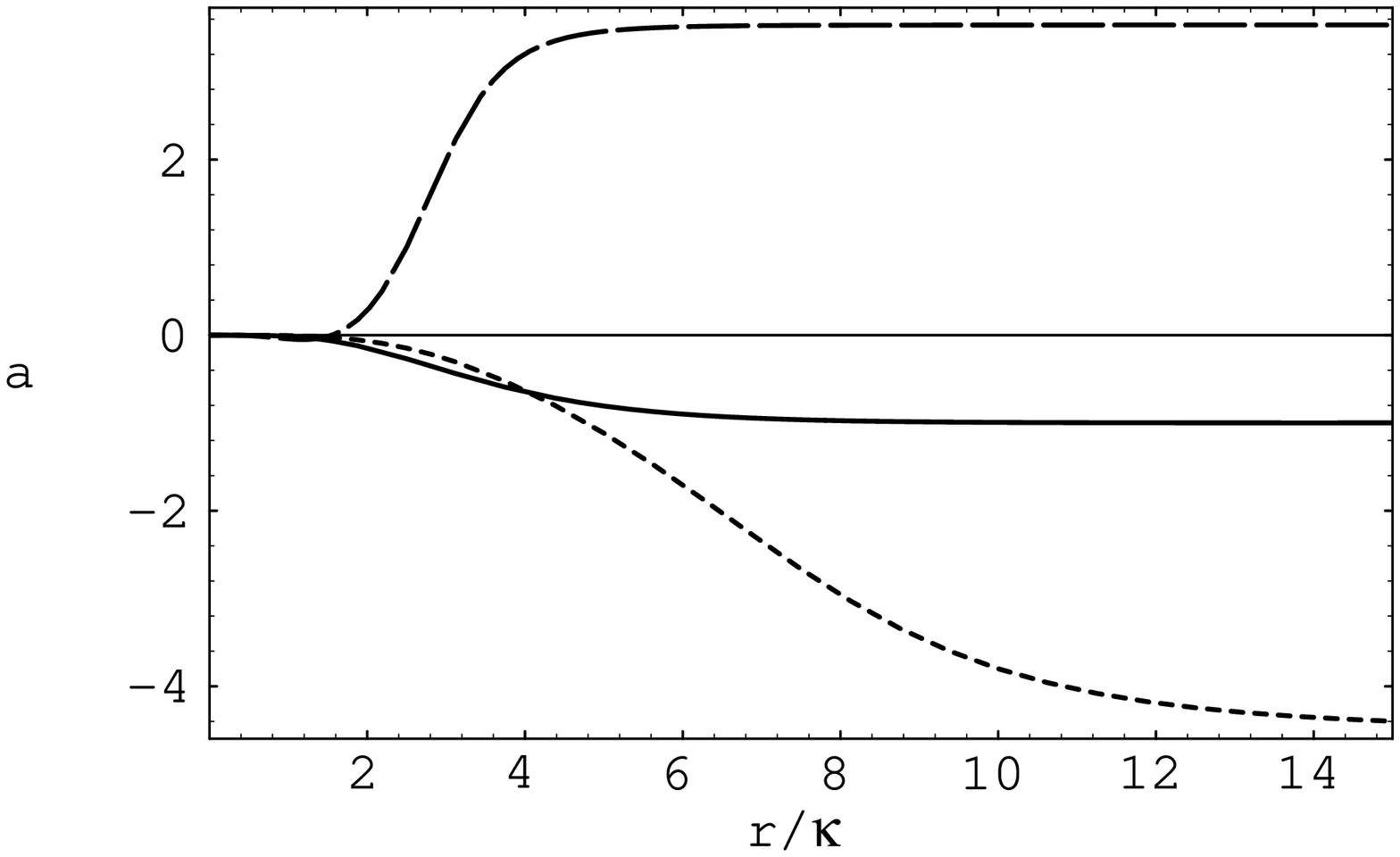}}
\hangindent\parindent{{\bf Fig.4}
Plotting $a(r)$ for $N=1$, $f(0)=0$ and $f_v=\pi/3$.  The dashed line
is for $\alpha=3.53325$, the solid line for $\alpha=-1.0$ and the
dotted line for $\alpha=-4.50025$ respectively.}

\vskip 2.0em

\noindent {\bf c)} When $f(0)=\pi $ and $N \le -1$.
\vskip1.5em

The solution in this case is similar to the previous one.  There are
again three kinds of solutions depending on $f(\infty) = 0 , f_v$ or
$\pi$. The behaviors of the function $f(r)$ and $a(r)$ are quite
similar to those in the previous case except their sign. Fig.5 shows
the function $f(r)$ for these three cases with $N=1,f_v=\pi/3$.  Fig.6
draws $a(r)$ for these cases.
\vskip 1.5em
\centerline{\epsfysize=5cm\epsffile{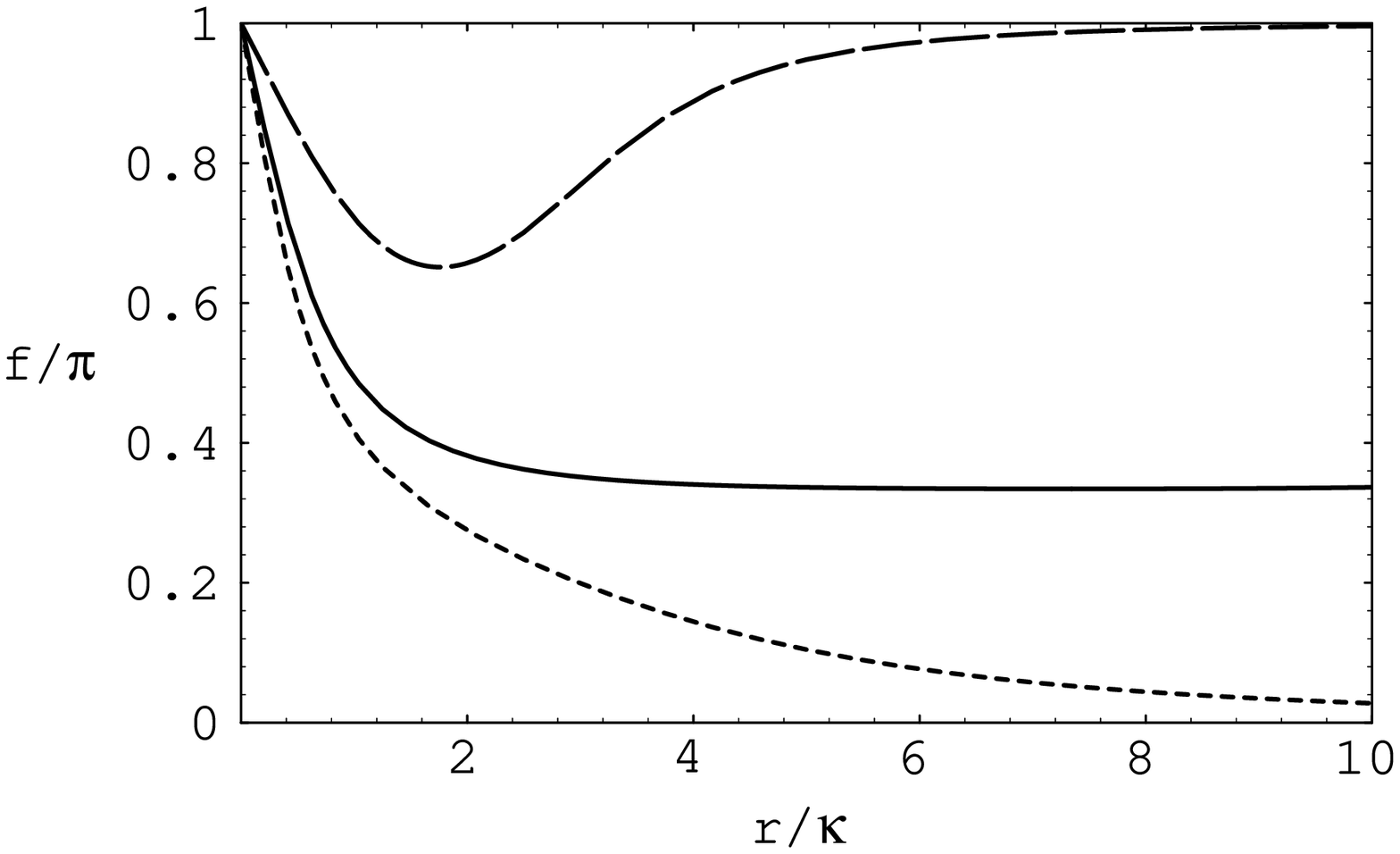}}
\hangindent\parindent{{\bf Fig.5}
Plotting $f(r)$ for $N=-1$, $f(0)=\pi$ and $f_v=\pi/3$.  The dashed
line is for $\alpha=4.77074$, the solid line for $\alpha=1.0$ and the
dotted line for $\alpha=-1.29324$ respectively.}
\vskip 1.5em
\centerline{\epsfysize=5cm\epsffile{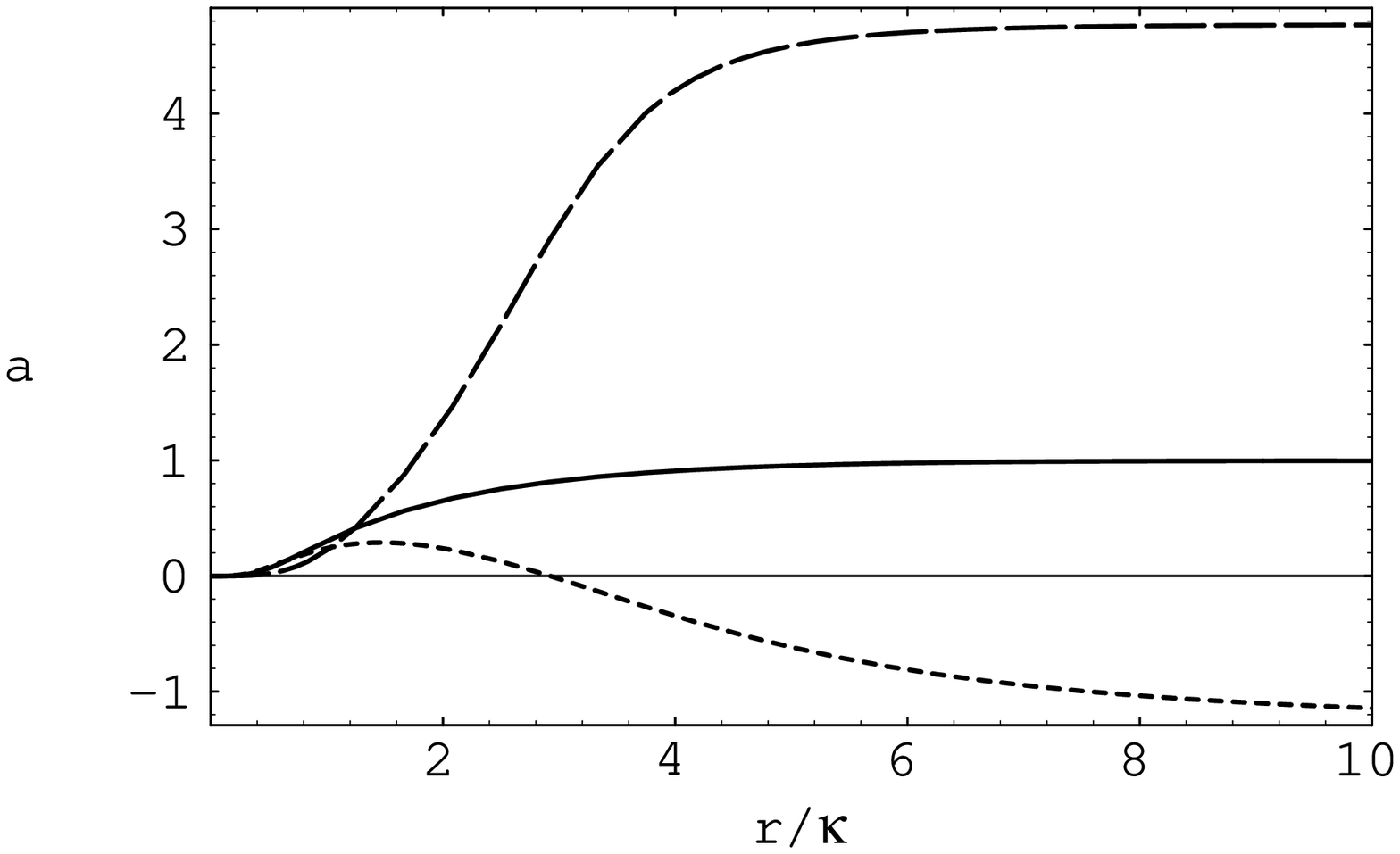}}
\hangindent\parindent{{\bf Fig.6}
Plotting $a(r)$ for $N=-1$, $f(r)=\pi$ and $f_v=\pi/3$.  The dashed
line is for $\alpha=4.77074$, the solid line for $\alpha=1.0$ and the
dotted line for $\alpha= -1.29324$.  }

\vskip2.0em

\noindent{\bf d)} When $|v| \ge 1$.
\vskip 1.5em

Noting that the case of $v>1$ becomes identical to the case of $v<-1$
if we exchanges $ f $ with $\pi-f$, let us focus ourselves on the case
$v<-1$. We could regard the solutions in this case as some continuous
deformation of the solutions in the case $|v|<1$.  Among the solitons
discussed in the previous sections, those solutions with $f(r)$ lying
between $\pi$ and $f_v$ will no longer exist.

When $N=0$, the nontopological solitons without vortices exist only
with $\pi>f(0)>0$ and $f(\infty)=0$. Both $f,a$ are decreasing
functions and $\alpha<-1$.  When $f(0)=0$ and $N\ge 1$, there are two
kinds of solutions depending on whether $f(\infty)=\pi$ or $0$.  For
$f(\infty)=\pi$, they are topological lumps. The function $f(r)$ is
increasing and $a(r)$ is decreasing to $\alpha$. It should be that
$\alpha>1/2-N$ if $v=-1$ and that $\alpha>1-N$ if $v<-1$.  For
$f(\infty)=0$, they are nontopological solitons with vortices in the
middle. The function $f(r)$ is increasing and then decreasing and
$a(r)$ is a decreasing function with $\alpha<-(1+N)$.  When $f(0)=\pi$
and $N \le -1 $, there are topological lumps with $f(\infty)=0$. Both
$f(r)$ and $a(r)$ are decreasing functions and $\alpha <0$.

We have seen that the various kinds of solitons in the symmetric and
asymmetric phases. In the symmetric phase there exist nontopological
solitons with or without vortices as well as topological ones whose
degrees are nonzero.  The magnetic flux of solitons are not in general
quantized. The mass and size of the solitons depend on their magnetic
fluxes except when $v=\pm 1$ and $f(\infty)= v$. They in general carry
the nonzero fractional angular momenta. In the asymmetric phase there
can be a topological vortex whose degree takes a fractional value. Its
magnetic flux is quantized and its mass is proportional to its
vorticity.

The topological vortices in the broken phase with $|v|<1$ can be
classified into two classes for a given positive $T$, depending on
whether $f(0)= 0$ or $ \pi $.  Their magnetic fluxes have opposite
signs, but their angular momenta are along the same direction.  We can
imagine a configuration containing these two different vortices, which
may be separated apart. As their topological charges $T$ add up, their
energy is bounded by the sum of their masses.  Interesting question is
then whether there is any self-dual configuration which represents
this composite state.  Such a self-dual configuration, if exits,
cannot have rotationally symmetric configuration because the points
where $f=0$ and the points $f=\pi$ cannot be arranged to be
rotationally symmetric.

We can extend our work in many directions. Clearly, we can introduce
the parameter $v$ in the pure Maxwell system, where we expect two
types of vortices in the broken phase as in the pure Chern-Simons.
There also can be topological lumps which are rotationally asymmetric.
The statistical phases of vortices with fractional spins were
explained by including the phase due to the Magnus force\cite{Klee}.
When there is a uniform external charge, due to the Magnus force
single vortex behaves like a charged particle in a uniform magnetic
field. It would be interesting to find out how those two types of
vortices interacting with each other.  The phase structure of the
Chern-Simons Higgs system with a uniform background charge was found
to be rich and interesting\cite{LeeYi}.  Our non-linear $O(3)$ models
with the background charge will have even richer structures and
deserve further investigation.

\vspace{1cm}
\acknowledgements{KK was supported in part by KOSEF through the Center for
Theoretical Physics of Seoul National University. KL was supported in
part by the NSF Presidential Young Investigator Program and the Alfred
P. Sloan Foundation. TL was supported by the Basic Science Research
Institute Program, Ministry of Education, 1995, Project BSRI-95-2401.}


\begin{thebibliography} {99}
%
\bibitem{gaugeo3-1} B.J. Schroers, Phys. Lett. B 356, 291 (1995).
%
\bibitem{gaugeo3-2} J. Gladikowski, B.M.A.G. Piette and B.J. Schroeres,
                    DTP 95-29 (hep-th/950699).
%
\bibitem{wilczek} F. Wilczek and A. Zee, Phys. Rev. Lett 51, 2250
(1983); M.J. Bowick, D. Karabali, and L.C.R. Wijewardhana, Nucl. Phys.
B 271, 417 (1986).
%
\bibitem{kara} D. Karabali and G. Murthy, Phys. Rev. D 35, 1522 (1987).
%
\bibitem{tlee} T. Lee, C. Rao, and K.S. Viswanathan, Phys. Rev. D 39,
2350 (1989).
%
\bibitem{Ghosh} P.K. Ghosh and S.K. Ghosh, IP/BBSR/95-54 (1995).
%
\bibitem{Leese}
R.A. Leese, Nucl.Phys. B344 (1990) 33; Nucl. Phys. B 366, 283 (1991).
%
\bibitem{belavin} A.A. Belavin and A.M. Polyakov, JETP Lett. 22 (1975) 245
%
%
\bibitem{Klee} K. Lee, Phys. Rev. D 49, 4265  (1994).
%
\bibitem{LeeYi} K. Lee and P. Yi, Phys. Rev. D 52, 2412 (1995).
%
\bibitem{Bogo} E. B. Bogomol'nyi, Yad. Fiz. 24, 861  (1976)
                [Sov. J. Nucl. Phys. 24 (1976) 449].
%
\bibitem{Hong} J. Hong, Y. Kim, P.Y. Pac, Phys. Rev. Lett 64, 2230
(1990); R. Jackiw and E.J. Weinberg, Phys. Rev. Lett 64, 2234 (1990).
%
%
\bibitem{jlw} R. Jackiw, K. Lee, and E. J. Weinberg, Phys. Rev. D 42, (1990)
               3499; Y. Kim and K. Lee, Phys. Rev, D 49. 2041 (1994).
%
\end{thebibliography}
\end{document}